\documentclass[journal,comsoc]{IEEEtran}
\usepackage[table]{xcolor}

\usepackage[T1]{fontenc}

\ifCLASSINFOpdf
\else
\fi

\usepackage{amsmath}
\usepackage{bm}
\usepackage[cmintegrals]{newtxmath}

\usepackage{balance}
\usepackage[caption=false]{subfig}
\usepackage{graphicx}
\usepackage{subfig}
\usepackage{verbatim}
\usepackage{array}
\usepackage{multirow}
\usepackage{float}
\usepackage{url}
\usepackage{tabularx}
\usepackage{tikz}
\usepackage{cite}
\usepackage[ruled,vlined]{algorithm2e}
\include{pythonlisting}
\colorlet{mygreen}{green!60!gray}

\usepackage{tablefootnote}
\usepackage{threeparttable}
\usepackage{hyperref} 
\begin{document}

\title{
Mitigating Label Flipping Attacks in Malicious URL Detectors Using Ensemble Trees}

\author{
Ehsan~Nowroozi, \IEEEmembership{Senior Member,~IEEE}, 
~Nada~Jadalla, ~\IEEEmembership{Member,~IEEE}, 
~Samaneh~Ghelichkhani, ~\IEEEmembership{Member,~IEEE}, 
~Alireza~Jolfaei, ~\IEEEmembership{Senior Member,~IEEE} 
\IEEEcompsocitemizethanks{
\IEEEcompsocthanksitem E. Nowroozi is with the Department of Business and Computing, Ravensbourne University London (RUL), United Kingdom (e-mail: e.nowroozi@rave.ac.uk) - Corresponding Author.
\IEEEcompsocthanksitem N. Jadalla is with the Bahcesehir University (BAU), Master in Cybersecurity, Istanbul, Turkey.
 (e-mail: nada.jadalla@bahcesehir.edu.tr)
\IEEEcompsocthanksitem S. Ghelichkhani is with the University of Leeds, Faculty of Engineering and Physical Sciences Master (Computing), Master in Advanced Computer Science, United Kingdom.
 (e-mail: samanehghelichkhani@gmail.com)
 \IEEEcompsocthanksitem A. Jolfaei is with the College of Science and Engineering at Flinders University, Adelaide, Australia. 
  (e-mail: alireza.jolfaei@flinders.edu.au)
}
}

\maketitle

\begin{abstract}

Malicious URLs provide adversarial opportunities across various industries, including transportation, healthcare, energy, and banking which could be detrimental to business operations. Consequently, the detection of these URLs is of crucial importance however, current Machine Learning (ML) models are susceptible to backdoor attacks. These attacks involve manipulating a small percentage of training data labels, such as Label Flipping (LF), which changes benign labels to malicious ones and vice versa. This manipulation results in misclassification and leads to incorrect model behavior. Therefore, integrating defense mechanisms into the architecture of ML models becomes an imperative consideration to fortify against potential attacks.

The focus of this study is on backdoor attacks in the context of URL detection using ensemble trees. By illuminating the motivations behind such attacks, highlighting the roles of attackers, and emphasizing the critical importance of effective defense strategies, this paper contributes to the ongoing efforts to fortify ML models against adversarial threats within the ML domain in network security. We propose an innovative alarm system that detects the presence of poisoned labels and a defense mechanism designed to uncover the original class labels with the aim of mitigating  backdoor attacks on ensemble tree classifiers.
We conducted a case study using the Alexa and Phishing Site URL datasets and showed that LF attacks can be addressed using our proposed defense mechanism.
Our experimental results prove that the LF attack achieved an Attack Success Rate (ASR) between 50-65\% within 2-5\%, and the innovative defense method successfully detected poisoned labels with an accuracy of up to 100\%.


\end{abstract}


\section{Introduction}
\label{sec:intro-section}

\IEEEPARstart{W}{hile} URLs (Uniform Resource Locators) play a crucial role in web browsing and the overall functioning of the World Wide Web, they have also been used as gateways to adversely impact and detrimentally affect users and businesses. To address this issue, extensive research has been conducted using ML systems, such as the Random Forest (RF) models \cite{nowroozi2022adversarial}. However, these models are susceptible to backdoor attacks due to their vulnerability to biased or manipulated training data.
RF is a widely adopted ensemble learning technique used across various domains throughout recent years due to its ability to handle complex datasets, feature interactions, and noisy data however, it can be vulnerable to adversarial attacks e.g. poisoning attacks. Malicious URLs can be poisoned in different ways to fool detection systems such as RF to evade detection and appear legitimate. One approach is through flipping the labels of the URLs from malicious to benign and benign to malicious with the aim to fool the system leading to malicious content for gaining unauthorized access or disrupting the systems. 
You can find several papers on the detection of malicious URLs in addition to different software deployed to block malicious URLs such as Symantec Endpoint Protection \cite{SymantecEndpointProtection}, McAfee \cite{McAfee}, Cisco Umbrella \cite{CiscoUmbrella}. However, the focus is more on the detection of malicious URLs, which is carried out mostly through the usage of common blacklists that include both benign and malicious URLs, \cite{sankaran2021detection} but not on the implementation of a URL defense strategy. Introducing a defense strategy is important and must be implemented in all ML systems since blacklisting cannot detect unknown or new malicious URLs and attackers are capable of modifying the URL's characteristics and tweaking their approach to avoid detection leading to a successful injection of poisoned data into the training dataset.


Attacks targeting ML systems are classified as either Poisoning (Causative) Attacks or Evasion (Exploratory) Attacks, based on the phase during which the attack is launched \cite{ahmed2021threats, Nowroozi2023Ensemple}. 
Poisoning attacks are launched during the training phase to manipulate, disrupt, or impact the ML system such as Clean Label Attacks, LF attacks, and Backdoor attacks \cite{lin2021ml}. While evasion attacks are applied during the testing phase with the aim to produce adversary-selected outputs without tampering with the ML model such as Confidence Score Attacks, Gradient Attacks, and Hard Label Attacks \cite{moisejevs2019evasion}. Furthermore, attacks can be classified into targeted attacks, where the attacker changes the behavior of the model on particular instances, or into untargeted attacks, where the attacker aims to impact a model’s performance with random instances or scenarios \cite{shahid2022label}. Several research studies have employed different strategies to implement poisoning attacks including LF attacks through different threat models and usually worst case scenarios are studied, where the attacker is capable of poisoning the training dataset directly aiming to impact the model's behavior. For instance, paper \cite{chang2020data} designed a data poisoning attack on an RF-based ML model with the aim of decreasing the model's classification accuracy, while paper \cite{dunn2020robustness} proposed poisoning attacks by using label modification functions on ML and Deep Learning (DL) models, along with varying poison rates, to measure performance degradation.  

Over the years, researchers in the ML field mainly focused on listing out existing detection mechanisms \cite{carlini2017adversarial} and how to enhance them \cite{gilmer1807motivating}, since targeted cyber-attacks are changing regularly with further enhanced attack strategies. Paper \cite{kumar2020adversarial} conducted a comprehensive survey to identify and fill the gaps in protecting ML systems within different organizations. This survey indicates that organizations are under threat from poisoning attacks more than other attacks however, the right security tools are not being used to protect their systems despite the importance of AI security to the operation of their business \cite{kumar2020adversarial}. The usage of ML systems is rapidly growing in the software industry however, organizations seem to have a lack of knowledge on securing their ML systems. Top Organizations like Google and Microsoft have called for initiatives to secure ML systems \cite{GoogleAdv}, \cite{MicrosoftAdv}, and several studies have been published on applying poisoning attacks and building defense mechanisms through different approaches such as K-LID \cite{shahid2022label}, PEFL \cite{liu2021privacy}, and so forth.  


The purpose behind our study is to investigate, highlight, and mitigate the vulnerabilities of RF classifiers introduced by LF attacks using corrupted training sets within the domain of ensemble tree-based URL classification and prove the efficiency of the proposed defense mechanism. Through comprehensive analysis of LF attack strategies and their impact on ensemble tree models, we developed an innovative defense method, which is the first mechanism able to identify malicious URLs from benign URLs by detecting manipulated labels, identifying their true label, and ultimately improving the robustness of RF-based malicious URL detection systems. By bridging the gap between backdoor attacks and ensemble tree classifiers, our work contributes to the field of ML security, advancing our understanding of effective countermeasures to defend against highly developed threats in real-world cybersecurity scenarios. Our simulation Python is publicly available at Github \cite{GitHub}.

\subsection{Contributions}
The contributions of our paper are briefly listed in the following points:
\begin{itemize}
  \item We initially ran the RF model on six clean datasets \cite{nowroozi2022adversarial} to observe the model's performance and detect any future behavioral changes. We obtained 99.87\% training accuracy for Dataset 6 and 100\% training accuracy for the other five datasets.
  \item We apply LF attacks on six RF models by applying different poison rates (2-5\%) on training datasets and analyzed the LF impact on the model’s behavior. The results show that the RF model was not able to detect the attack and the attack successfully fooled the RF model with an ASR higher than 50\%. For Dataset 5\cite{nowroozi2022adversarial}, we obtained 57.14\% ASR with a 4\% poison rate and 61.09\% ASR with a 5\% poison rate. 
  \item We evaluate the effectiveness of applying the \textit{K}-NN method with the aim of detecting poisoned labels and predicting their true label. Then ran the RF model with the recovered datasets and observed an increase in training accuracy. For instance, training accuracy increased from 96.08\% to 99.89\% for Dataset 5 recovered from 4\% poisoned data,  and training accuracy increased from 95.19\% to 100\% for Dataset 5 recovered from 5\% poisoned data.
\end{itemize}

\subsection{Organization}
This paper consists of five sections. Section \ref{sec:intro-section} presents the introduction, which covers the main contributions, motivations, and rationale of our research. Our primary objective is to propose a detection and defense method against adversarial attacks, in specific, LF attacks that target RF classifiers. 
The remaining sections are organized as follows: Section \ref{sec:related-section} presents the related work within our domain. Section \ref{sec:meth-section} explains the applied methodology in our study including the used datasets, selected classifier, applied LF-based attack, and proposed defense method against such attacks. Section \ref{sec:results-section} presents the results of running clean datasets, poisoned datasets, and recovered datasets. Finally, Section \ref{sec:conc-section} concludes our research and discusses future work.


\section{Related Work}
\label{Related}
\label{sec:related-section}

Due to the increase in digitization and high usage of technologies, ML models are susceptible to poisoning attacks causing it to be an emerging research topic over the recent years \cite{goldblum2022dataset}. For instance, paper \cite{shahid2022label} designed a white-box attack on a Human Activity Recognition (HAR) by randomly altering the labels from the training dataset using different algorithms including RF, Multi-layer Perceptron (MLP), XGboost, and Decision Tree (DT). This attack has minimized the loss function on true data however, the system may notice some changes in data which limits the attacker's capability to further cause larger damage to the model. While paper \cite{ramirez2022new} applied both untargeted and targeted LF attacks on five different ML models including DT, and RF models. Both attacks have caused a degradation in the accuracy of all ML classifiers and increased the misclassification rate however, the attacker must access the dataset during the training phase to be able to perform the attack. Another study proposed poisoning attacks against IoT fake packet classifiers by using label modification functions on RF-based ML models using different poison rates to illustrate the degradation in the model’s performance\cite{dunn2020robustness}. In paper \cite{yerlikaya2022data}, a random LF attack is designed on different ML models, including the RF model, with the aim to cause an impact on the accuracy of the models. A minor drop in accuracy of RF was observed within 0-12.5\% poisoning rate yet a further increase in poison rate has caused a higher degradation in accuracy rate which means that the model has detected the poisoned labels.


\begin{table*}[ht!]
\centering
\renewcommand{\arraystretch}{0.9}
\caption{Comparison between previous works and our's}
\label{table:relatedworks}
\resizebox{\linewidth}{!}{%
\begin{tabular}{|p{0.5cm}|p{1.8cm}|p{3.0cm}|p{6.0cm}|p{5.0cm}|}
\hline
\textbf{Ref.} & \textbf{ML Model} & \textbf{Utilized Datasets}  & \textbf{Advantages} & \textbf{Disadvantages} \\
\hline
\cite{gardiner2016security} & Different classifiers & 10 real-world datasets & - Detects sophisticated attacks   & - Requires traffic storage \\ &&& - Reduces false positive & \\
\hline
\cite{paudice2019label} & Linear Classifier & MNIST\cite{deng2012mnist}, Spambase\cite{misc_spambase_94} and BreastCancer\cite{misc_breast_cancer_wisconsin_(diagnostic)_17} & - Effective Against Label Flipping Attacks &  - Sensitivity to parameters \\ &&& - Applicability in Various Scenarios & - Scalability \\
\hline
\cite{paudice2018detection} & Linear Classifier & MNIST\cite{deng2012mnist}, Spambase\cite{misc_spambase_94} and BreastCancer\cite{misc_breast_cancer_wisconsin_(diagnostic)_17} & - Detects attack points and Outlier elimination & - Computationally intensive and requires outlier estimation \\
\hline
\cite{chan2021causative} & Ensemble Trees & UCI ML Repository\cite{UCIDataset} \& KEEL-dataset Repository\cite{alcal2011keel} & - Achieves high detection accuracy and handles multiple attack types & - Does not locate attacked points and requires untainted data for training \\
\hline
\cite{shahid2022label} & Ensemble Trees & HAR dataset\cite{misc_human_activity_recognition_using_smartphones_240} & - Recovers poisoned data, and increases accuracy & - Limited effectiveness   \\
\hline
\cite{anisetti2022robustness} & RF & Musk2\cite{misc_musk_(version_2)_75} and Android malware & - Acheived high accuracy under other perturbations and scalable approach  & - Ensemble size must be considered and dependent on adversary knowledge \\
\hline
\cite{cheng2021data} & AdaBoost & Spambase\cite{misc_spambase_94}, Breast-w\cite{misc_breast_cancer_wisconsin_(original)_15}, Kr-vs-kp\cite{misc_chess_(king-rook_vs._king-pawn)_22}, and so forth & - Combines weak classifiers into a strong classifier & - Sensitive to noisy \& abnormal data \\
\hline
\cite{tavallali2022adversarial} & AdaBoost & MNIST\cite{deng2012mnist} & - Applicable across various ML models \&  detect flipped labels during the training process  & - Learning problem \& label flipping budget constraint \\
\hline
\cite{taheri2020defending} & CNN & Drebin\cite{arp2014drebin}, Contagio\cite{ContagioDataset}, and Genome\cite{zhou2012dissecting}  & -Addresses IoT malware detection & - Proposing innovative method, called Silhouette clustering-based attack  \\ &&& - limit to other platforms \& defense methodology might not generalize to unseen malware samples. & \\
\hline
Our's & RF & URL Datasets in Table \ref{table:datasets}  & - High detection accuracy of poisoned labels (See Table\ref{table:Defense-on-all}), and considering attack and defense in Dataset URLs for the first time in computer networks (see Table\ref{table:BMdatasets} and \ref{table:datasets}) & - Future work: Difficult to apply feature positioning attack with compared to label poisoning, since the datasets consist of numerical and lexical features.  \\
\hline
\end{tabular}
}
\end{table*}

The importance of developing a model-agnostic defense has been today's research topic to detect and defend against poisoning attacks targeting different ML models. Paper \cite{gardiner2016security} introduced several Command \& Control (C\&C) detection systems to detect sophisticated attacks and reduce false positive rates, however, as network size increases, it is a challenging task to store all network traffic yet, it is an essential requirement of most C\&C detection methods to store traffic data. While in paper \cite{paudice2019label}, the aim is to detect poisoned labels through applying the \textit{K}-NN approach and mitigate the impact of LF attacks through label sanitization, however, it assumes a large number of benign samples are available for sanitization, which may reduce classifier accuracy. Another paper suggested to use of an outlier detection-based scheme to identify attack points targeting linear classifiers \cite{paudice2018detection}. However, the application of this scheme can be challenging in large high-dimensional datasets. On the other hand, paper \cite{chan2021causative} presents how the detection of causative attacks can enhance learning robustness through a two-step secure classification model that uses data complexity measures, yet, these measures calculate the difficulty of classification through computing the geometrical characteristics of data. Furthermore, paper \cite{shahid2022label} discusses the usage of \textit{K}-NN defense scheme to predict the true label and recover most of the data leading to a sharp increase in the model’s accuracy. Paper \cite{anisetti2022robustness} proposes a hash-based ensemble approach with the aim of increasing the robustness of RF models taking into account that the size of the ensemble is essential to avoid a drop in a model’s accuracy due to the usage of a large ensemble considered to be oversized. Paper \cite{taheri2020defending} presents a Label-based Semi-supervised Defense (LSD) that works by finding the poisoned samples and a Clustering-based Semi-supervised Defense (CSD) that uses clustering techniques to recover true labels. LSD and CSD increase the model’s accuracy rate, however, they are considered to have poor performance with reference to other defense methods in terms of speed. In paper \cite{cheng2021data}, a boosting ensemble method for two classifications is used through training a number of weak classifiers using the same training dataset for classifiers to merge into a much stronger classifier. However, noisy and abnormal data can easily impact the model due to its high sensitivity to such data. Paper \cite{tavallali2022adversarial} applies the Regularized Synthetic Reduced Nearest Neighbor (RSRNN) defense method by checking if any of the samples are above the confidence range, then this sample will be considered malicious. RSRNN was able to achieve the smallest test error while also detecting a high portion of malicious samples, however, the problem in learning an SRNN model is similar to that of a \textit{K}-means problem which is identifying the most relevant \textit{K} value of samples that will provide the lowest error rate. Table \ref{table:relatedworks} summarizes different detection and defense methods to protect ML models from LF attacks.

\section{Methodology}
\label{sec:meth-section}
In this section, we address the considered datasets, possible LF, and defense methodology. 

\subsection{Datasets}
This paper used a dataset of 3,980,870 benign and malicious URLs \cite{nowroozi2022adversarial}, gathered from twelve different datasets. Then six datasets are formed by merging the six benign URL datasets and the six malicious URL datasets in Table \ref{table:BMdatasets}.

\begin{table}[H]
\centering
\caption{The Selected Benign and Malicious Datasets \cite{nowroozi2022adversarial}}
\label{table:BMdatasets}
\resizebox{\linewidth}{!}{%
\begin{tabular}{|l|l|l|} \hline
\textbf{No.} & \textbf{Benign Datasets} & \textbf{Malicious Datasets}\\ \hline
1~ & Pristine Alexa & Phishing Site URL \\ \hline
2~ & Pristine Crowdflower & Phishtank  \\ \hline
3~ & Pristine DMOZ & Malicious data URL    \\ \hline
4~ & Benign set URL & ISCX-URL-2016  \\ \hline
5~ & Non-malicious URL & Phishstrom \\ \hline
6  & Pristine ISCX & Malicious set URL   \\ \hline
\end{tabular}
}
\end{table}

Table \ref{table:datasets} shows the six merged datasets including both benign and malicious URLs labeled as ’0’ and ’1’, respectively.
To further analyze the datasets, a statistical study identified that malicious URLs within the selected six datasets present a wider range of variability in their length compared to URLs that are considered benign in \cite{nowroozi2022adversarial}. However, in our poisoning scenario, we are targeting the labels of both benign and malicious URLs.

\begin{table}[h!]
\centering
\caption{The combined datasets were obtained from \cite{nowroozi2022adversarial}}
\label{table:datasets}
\resizebox{\linewidth}{!}{%
\begin{tabular}{|l|l|} \hline
\textbf{Dataset Name} & \textbf{Combined Datasets }                          \\ \hline
Dataset 1~ & Pristine Alexa and Phishing Site URL                   \\ \hline
Dataset 2~ & Pristine Crowdflower and malicious Phishtank                     \\ \hline
Dataset 3~ & Pristine DMOZ and Malicious data URL                   \\ \hline
Dataset 4~ & Benign set URL and malicious ISCX  \\ \hline
Dataset 5~ & Non-malicious URL and malicious Phishstrom              \\ \hline
Dataset 6 & Pristine ISCX and Malicious set URL  \\ \hline
\end{tabular}
}
\end{table}

In order to achieve accurate results, it is necessary to adjust and restructure the dataset to align with the input format needed for our dependencies. During the preprocessing phase, we apply the interquartile range technique to rescale the data. All duplicates and unprocessed cells containing values are eliminated, along with excluding repetitive hostname URLs. Subsequently, the collections of URLs are mixed up. We extract samples from these datasets for more in-depth analysis. Accordingly, datasets are ready to be used with the ML e.g. RF.

\subsection{Classifier}
In this study, RF model is the selected classifier which is a type of supervised learning based on using several decision trees to confirm a single output. The foundation of RF lies in the idea that while each individual tree might make accurate predictions, it is highly likely that some trees will excessively tailor themselves to specific data points, leading to overfitting. To counter this issue, the approach involves combining the outcomes of numerous trees, each excelling and overfitting in distinct ways, which effectively mitigates overfitting. This reduction in overfitting is achieved without compromising the predictive capability of the trees. The key parameters that can be adjusted include $n$ estimators, max features, and pre-pruning settings like max depth. For $n$ estimators, a larger value is recommended to ensure better performance. With an increased number of trees being averaged, the ensemble becomes more resilient by curbing overfitting. However, it is important to note that employing more trees also demands greater memory and training time. The mathematical representation of an RF can be expressed as follows:

Let \(D\) represent the original training dataset, and \(T\) denote the number of decision trees in the ensemble. For each tree \(t\), a random subset \(D_t\) of the training data, \(D\) is drawn, typically by bootstrapping (sampling with replacement). Additionally, a random subset of features \(F_t\) is selected for each node of the tree, which is a subset of the total features \(F\) available in the dataset. The decision tree \(t\) is then constructed using the data \(D_t\) and features \(F_t\) based on a specified criterion, such as Gini impurity or information gain.

During prediction, each tree \(t\) in the ensemble independently classifies or predicts a target variable. For classification tasks, the class, $y_c$, is determined by a majority vote among the trees (mode of class predictions) and can be represented mathematically as:

\begin{equation}
    \begin{split}
	{y_c} & = \text{Mode}(y_t) \text{ for } t = 1, 2, \ldots, T
    \end{split}
\end{equation}


while for regression tasks, the predicted value ,$y_r$, is obtained by averaging the predictions of all trees and can be represented mathematically as:

\begin{equation}
    \begin{split}
	{y_r} & = \frac{1}{T} \sum_{t=1}^{T} y_t
    \end{split}
\end{equation}


The aggregation of predictions from multiple randomized decision trees	helps minimize the overfitting issue and enhances the performance of the model. RF is a powerful and widely used ML due to its robustness and effectiveness, making it applicable in various domains, including classification, regression, and feature importance analysis.

Since our six datasets are a combination of numerical and lexical features, therefore, only decision trees have the ability to handle multiclass classification. Accordingly, RF is selected to be used in our study.

\subsection{LF Based Attack}

A random LF attack is applied to the six datasets with the aim of manipulating them by altering both benign and malicious labels randomly. Random LF attacks are a form of adversarial attack, where the attacker seeks to exploit vulnerabilities in the ML by manipulating the input data. 

\textit{\textbf{Black box}, \textbf{gray box}, and \textbf{white box}} attacks are the three main types of attacks of adversarial attacks. In a \textit{black box} attack, the attacker has no access to the system and this is usually considered the worst-case scenario in terms of system accessibility. In a \textit{white box} attack, the attacker has full access to the classifier including its weights and parameters. In this paper, we applied a \textit{gray box} attack, where the attacker has limited access to the classifier during a training phase. The threat model of such attacks is related to the attacker's goal, the attacker's knowledge, and the the attacker's capability to impact the training dataset \cite{wang2022poisoning}.
\paragraph{\textbf{Attacker's goal}} clarifies the kind of security violation and what error the attacker aims to apply. In this paper, the goal is to manipulate the training dataset used by the RF model and fool the model by concealing the poisoned labels.
\paragraph{\textbf{Attacker's knowledge}} the level of knowledge is determined based on the attacker’s access to the feature space, the target classifier, the model parameters, and the training dataset. The attacker in this paper has only access to the training dataset thus, the attacker's knowledge is considered as limited knowledge.
\paragraph{\textbf{Attacker's capability}} This is defined as the degree of control that the adversary possesses over both the training and testing data, as stated in \cite{nowroozi2022adversarial}.

In this paper, we split the six datasets into 79\% for training set and 21\% for testing set. Then, applied random LF attack by altering both benign and malicious labels randomly by considering five scenarios: 2\%, 3\%, 4\%, and 5\% poisoning rate.  

Our attack method is presented in the following Algorithm. As an input, this algorithm takes the RF, original dataset \textit{D}, and a number of poisoned samples based on the poisoning rate.

 


\begin{itemize}
    \item \textbf{Input:} Original Dataset $\mathcal{D}$, Poisoned Samples $X$.
    \item \textbf{Output:} Poisoned Dataset $\mathcal{D'}$, Accuracy Rate.
    \item For each sample $x_i$ in $X$:
    \begin{itemize}
        \item Select $X'$ as random rows from $\mathcal{D}$.
        \item Apply LF (a function or operation) on each $x_i$ in $X'$.
        \item Update the poisoned dataset $\mathcal{D}'$ with $x_i$.
        \item Train a model on the poisoned dataset $\mathcal{D}'$.
        \item Measure and record the accuracy of the model.
    \end{itemize}
\end{itemize}

Based on the selected poison rate, $X$ rows are selected randomly from $D$ to flip their labels. Let $D = \{(x_i, y_i)\}_{i=1}^N$ represent the original training dataset, where $x_i$ denotes the input data instances, $y_i$ denotes their corresponding true labels, and $N$ is the total number of data points. In a random LF attack, an attacker randomly alters a subset of the true labels as $y'_i$ which is the poisoned label and stored into a poisoned dataset, typically denoted as $D' = \{(x_i, y_i')\}_{i=1}^M$, where $M$ represents the number of instances that label manipulated. The manipulation involves changing randomly $y_i$ to $y_i'$, causing the model to learn the potentially erroneous labels.

LF attack is now applied by flipping the label randomly on each $X$ selected row in $D$. Labels denoted as "0" are flipped to "1" and treated as malicious samples in the training, and labels denoted as "1" are flipped to "0" and treated as benign samples. After executing the random LF process, the flipped labels are stored into $D'$. The RF is trained using $D'$ then the accuracy of both testing and training is recorded. The attack is triggered by testing the stored model with the labels of the entire flipped testing dataset, which measures the accuracy of the testing. The accuracy rate reported in this scenario reflects the ASR of labels that alter from an adversary of the intended attack approach.

\subsection{Proposed Defense Strategy}

The detection and defense method introduced in this paper aims to detect and sanitize the dataset from LF attacks. It works by taking original datasets, $D$, and untrusted datasets, $D'$, as input. Here we used the \textit{K}-NN method similar to the approach applied in \cite{paudice2019label} to detect flipped labels. This method trains a model using the inputs to predict the true label of a given URL sample and checks if it matches the current label of the URL sample. Our defense strategy is built as follows:

\paragraph{\textbf{Choose the Best $K$ Value}}


The selection of $K$ is crucial to avoid overfitting or underfitting the model. The value of $K$ determines the number of labels that need to be checked and helps in predicting the value of the tested label.

 

%
  %
   %
    %
     %

\begin{itemize}
    \item \textbf{Input:} Original Dataset $\mathcal{D}$, Untrusted Dataset $\mathcal{D}'$, $\mathcal{X}$ values of $K$.
    \item \textbf{Output:} $K$.
    \item For each $K$ in $\mathcal{X}$:
    \begin{itemize}
        \item For each $i$:
        \begin{itemize}
            \item Calculate the distance $d$ as the Euclidean distance between points in datasets $\mathcal{D}$ and $\mathcal{D}'$:
            \[ d = \sqrt{\sum_{i=1}^{n} (x_{i} - y_{i})^2} \]
            \item Determine the mode of the $K$ nearest labels:
            \[ \text{Mode} = \arg \max_i \left( \sum_{j=1}^{K} I(x_j = i) \right) \]
            \item If Mode equals 1, set $x_p$ to 1; otherwise, set $x_p$ to 0.
        \end{itemize}
        \item If $L_m = 0$, select $K$.
    \end{itemize}
\end{itemize}

This Algorithm determines the best value of \textit{K} within $X$ set of $K$ values and selects $K$ with respect to the minimum error rate. This algorithm takes an original dataset $D$, untrusted (poisoned) dataset $D'$, and $X$ set of $K$ values as inputs. In our study, we considered the following $X$ set of $K$ values: $X= \{1,3,5,7,9,\dots,33,35,37, \text{ and } 39\}$. For each $K$ value within $X$ set, the distance $d$ of labels in $D$ from the tested label in $D'$ is computed and the mode of $K$ nearest labels is found to confirm the label value. The most relevant $K$ value is selected when a number of mismatch labels, $L_m$, in $D$ equals to zero to minimize the error rate.

\paragraph{\textbf{Defense Method}}

In this study, we considered $K$-NN as our defense method to mitigate the impact of random LF attack. The $K$-NN algorithm can be expressed as follows:

Given a dataset \(D\) with data points represented as vectors in a \(d\)-dimensional feature space, let \(x_i\) and \(y_i\) be the target data points for which we want to make a prediction or classification and $n$ is the total number of labels in the dataset. The \(K\)-NN algorithm estimates distances between \(x_i\) and all other data points in \(D\), typically using metrics like the Euclidean distance or Manhattan distance defined as the following:

\begin{equation}
    \begin{split}
	d& = \sqrt{\sum_{i=1}^{n} \left( x_{i}-y_{i}\right)^2 }
    \end{split}
\end{equation}


Our detection and defense method is presented in the following Algorithm. This algorithm takes an original dataset $D$, an untrusted dataset $D'$, and a value of $K$ as inputs.

%
  %
 %
\begin{itemize}
    \item \textbf{Input:} Parameter \textit{K}, Original Dataset \textit{D}, Untrusted Dataset \textit{D'}.
    \item \textbf{Output:} True labels $x_i$, Alarm.
    \item For each index \textit{i}:
    \begin{itemize}
        \item Calculate the distance $d$ as the Euclidean distance between points in datasets \textit{D} and \textit{D'}:
        \[ d = \sqrt{\sum_{i=1}^{n} (x_{i} - y_{i})^2} \]
        \item Based on $d$, pick \textit{K} nearest labels from \textit{D}.
        \item Determine the mode of the \textit{K} nearest labels:
        \[ \text{Mode} = \arg \max_i \left( \sum_{j=1}^{K} I(x_j = i) \right) \]
        \item If Mode equals 1, set $x_p$ to 1; otherwise, set $x_p$ to 0.
        \item If $x_p$ is not equal to $x_i$:
        \begin{itemize}
            \item Raise an alarm indicating a potential LF Attack.
            \item Update $x_i$ to $x_p$.
        \end{itemize}
    \end{itemize}
\end{itemize}
      
 

After selecting the value of $K$ using Algorithm \ref{alg:K}, this method measures the distance $d$ of each label $y_i$ in the original dataset with reference to the input sample $x_i$ and then selects the $K$ nearest samples. The algorithm finds the mode of $K$ nearest labels and checks if mode equals 1 then the predicted label, $x_p$ is equal to 1 otherwise if mode equals 0 then, $x_p$ equals 0. Afterward, $x_p$ is compared with the value of $x_i$ and if $x_p$ is not equal to $x_i$, the system detects the LF attack, activates an alarm to notify the user and recovers the true label by storing $x_p$ into $x_i$.

The predicted value of the label is identified based on the mode of the $K$ nearest labels. To find the mode, sort the numbers from lowest to highest and observe which label is the most frequently appearing. For example, if the mode of $K$ nearest labels equals to ``1'' then the predicted label will equal to ``1'' and if the mode equals to ``0'', then the predicted label will equal to ``0''. Finally, the model checks if the predicted label matches the value of the label in the untrusted dataset $D'$. In case a mismatch is found, it flips the label to recover the true label and clean the dataset from LF attack. This process will be repeated until all labels in the dataset are checked and if any poisoned label is detected, the label is corrected and an alarm is activated to inform the user about the LF attack. Algorithm \ref{alg:KNN} describes the steps of the applied defense mechanism.


\section{Results and Discussion}
\label{Results}
\label{sec:results-section}

In our experiment, we utilized the six different datasets described in Table \ref{table:datasets} \cite{nowroozi2022adversarial}. Each dataset consists of 1000 URLs which involves both benign or malicious samples. Benign URLs have a label of ``0'' while malicious URLs have a label of ``1''. Different poison rates of LF attack are applied to each dataset through flipping both benign and malicious labels. We ran the original dataset $D$ on RF without any attack. The confusion matrix utilized in our investigation comprised four distinct values, namely: True Positive (TP), True Negative (TN), False Positive (FP), and False Negative (FN), which facilitated the visualization of the RF performance. Based on the confusion matrix, several performance metrics are computed by utilizing specific formulas to provide insights into the efficacy of the RF. The following set of metrics was used to calculate accuracy:

\textbf{A True Positive Rate (TPR)}, refers to positive URL samples classified as a true class.
\begin{equation}
    \begin{split}
	TPR & = \frac{TP}{TP+FN}
    \end{split}
\end{equation}

\textbf{A True Negative Rate (TNR)}, refers to positive URL samples that yield to negative test results.
\begin{equation}
    \begin{split}
	TNR & = \frac{TN}{TN+FP}
    \end{split}
\end{equation}

\textbf{A False Positive Rate (FPR)}, refers to all negative URLs that RF classifier generates it as positive samples.
\begin{equation}
    \begin{split}
	FPR & = \frac{FP}{FP+TN}
    \end{split}
\end{equation}

\textbf{A False Negative Rate (FNR)}, refers to all positive URLs samples that RF classifier generates it as negative samples.
\begin{equation}
    \begin{split}
	FNR & = \frac{FN}{FN+TP}
    \end{split}
\end{equation}

\textbf{Accuracy} is determined by the total number of examples that can be predicted with high reliability, relative to all examples in the six dataset.
\begin{equation}
    \begin{split}
	Accuracy & = \frac{TP+TN}{TP+TN+FP+FN}
    \end{split}
\end{equation}

Beside computing the accuracy for each dataset, we asses the performance of the adversary and RF by computing ASR.
ASR is calculated by measuring the model's accuracy when tested on a poisoned dataset. 
A detailed report of the obtained results for each scenario is presented in this section to evaluate both the model's performance and the impact of the utilized attack in addition to the effectiveness of our proposed defense method against random LF attacks.

\subsection{Run RF with Original dataset}

To establish a baseline mode for evaluating the attack and defense approach, we initially trained the RF model on original datasets to observe the accuracy behavior when no attack is applied on all six datasets in Table \ref{table:datasets}. The obtained training and testing accuracies on RF using original datasets are reported in Table \ref{table:cleanrun}. We can clearly observe from the results when no attack applied to RF model, no drop in accuracy for five datasets except Dataset 6 observed a minor drop that caused the accuracy rate to equal 99.87\% however this drop is considered negligible in ML systems. Taking into account that the acceptable range of accuracy drop can vary significantly based on the application, domain, and potential consequences of errors. 

\begin{table}[H]
\centering
\caption{Accuracy of RF model on the six datasets without attack}
\label{table:cleanrun}
\tiny 
\resizebox{\linewidth}{!}{%
\begin{tabular}{|l|l|l|} 
\hline
Dataset & Tr. Accuracy & Te. Accuracy \\ 
\hline
\multirow{1}{*}{\textbf{Dataset 1 ... 5}} & 100.00\% & 100.00\% \\ 
\hline
\multirow{1}{*}{\textbf{Dataset 6}} & 99.87\% & 100.00\% \\ 
\hline
\end{tabular}
}
\end{table}

Considering the original dataset on RF models can be used as a reliable reference point for both the attack and defense scenarios to evaluate model performance. Our aim is to propose a defense mechanism against random LF attacks that can detect manipulated URL labels and enhance the robustness of the RF model.

\begin{figure}[H]
    \centering
    \includegraphics[width = 0.9\columnwidth]{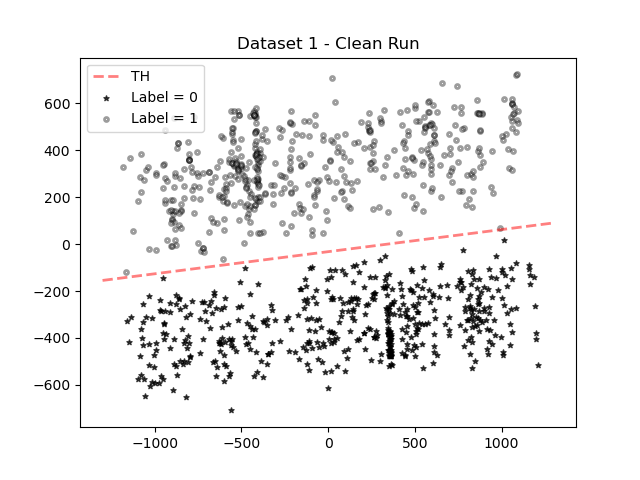}
    \caption{Distribution of Benign and Malicious labels from original Dataset 1}
    \label{fig:CleanRun Dataset 1}
\end{figure}

As shown in Figure \ref{fig:CleanRun Dataset 1}, Dataset 1 stands as an example of a clean dataset where benign URLs are labeled as ``0'', and malicious URLs are labeled as ``1'', both are separate from the decision margin. The red threshold acts as a decision margin that splits the benign samples from the malicious samples and helps to detect any manipulated label. 

\subsection{Apply LF attack on RF}

By considering Algorithm \ref{alg:RLF Attack}, we poisoned each dataset with 2\% up to 5\% causing 16 to 40 URL labels to be poisoned. The results of our random LF attack are reported in Table \ref{table:Randdom-on-all}.

Based on the results, we can understand that the utilized attack works since the attack barely had a 2\% effect on the RF model when 2\% of the data is poisoned. Even with introducing a minor increase in the poison rate (from 2\% to 5\%), the RF model's accuracy drop was restricted to no more than 5\% and the impact would not be considered significant since it did not influence the model's accuracy crucially. This approach maintained the attack's covert nature while permitting poisoned backdoor samples to be integrated into the training dataset.
The success of the attack was determined by calculating the ASR using all of the testing samples, which revealed the model's classification error rate. Normal testing accuracy measures the correct classification, while ASR measures the classification error. The ASR values surpassed 40\%, indicating that this method was successful in bypassing detection by the URL detector and served as a ghost attack, remaining unnoticed in the framework, in line with the study's threshold. For example in Table \ref{table:Randdom-on-all}, Dataset 1 obtained 98.35\% training accuracy and 57.62\% ASR with 2\% poisoning rate while Dataset 2 obtained 97.97\% training accuracy and 58.1\% ASR with 2\% poisoning rate. Both results present the attacker's success in manipulating Dataset 1 and Dataset 2 which led to fooling the RF model with only 2\% poisoned data.

\begin{table}[H]
\centering
\caption{Accuracy of RF model on the six datasets with Random LF attack}
\label{table:Randdom-on-all}
\resizebox{\linewidth}{!}{%
\begin{tabular}{|l|l|l|l|l|} 
\hline
Dataset & Poison\% & Poisoned Count & Tr. Accuracy~~ & ASR \\ \hline
\multirow{5}{*}{\textbf{Dataset 1}} & 2\% & 16 & 98.35\% & 57.62\% \\        
                           \cline{2-5}
                           & 3\% & 24 & 97.09\% & 50.47\% \\ \cline{2-5}
                           & 4\% & 32 & 95.95\% & 61.09\% \\ \cline{2-5}
                           & 5\% & 40 & 97.34\% & 55.23\%  \\ \hline
\multirow{5}{*}{\textbf{Dataset 2}} & 2\% & 16 & 97.97\% & 58.1\%  \\ 
                           \cline{2-5}
                           & 3\% & 24 & 96.96\% & 55.23\%  \\ \cline{2-5}
                           & 4\% & 32 & 96.08\% & 50.47\%  \\ \cline{2-5}
                           & 5\% & 40 & 95.18\% & 64.76\%  \\ \hline
\multirow{5}{*}{\textbf{Dataset 3}} & 2\% & 16 & 98.23\% & 50.48\%  \\ 
                           \cline{2-5}
                           & 3\% & 24 & 97.22\% & 53.33\%  \\ \cline{2-5}
                           & 4\% & 32 & 96.2\%  & 56.19\%  \\ \cline{2-5}
                           & 5\% & 40 & 95.19\% & 55.23\%   \\ \hline
\multirow{5}{*}{\textbf{Dataset 4}} & 2\% & 16 & 97.97\% & 55.24\%   \\  
                           \cline{2-5}
                           & 3\% & 24 & 97.47\% & 54.28\%   \\ \cline{2-5}
                           & 4\% & 32 & 96.33\% & 58.09\%   \\ \cline{2-5}
                           & 5\% & 40 & 95.19\% & 56.19\%   \\ \hline
\multirow{5}{*}{\textbf{Dataset 5}} & 2\% & 16 & 97.97\% & 62.38\%   \\ 
                           \cline{2-5}
                           & 3\% & 24 & 96.96\% & 53.33\%    \\ \cline{2-5}
                           & 4\% & 32 & 96.08\% & 57.14\%    \\ \cline{2-5}
                           & 5\% & 40 & 95.19\% & 61.09\%    \\ \hline
\multirow{5}{*}{\textbf{Dataset 6}} & 2\% & 16 & 98.10\% & 59.05\%    \\ 
                           \cline{2-5}
                           & 3\% & 24 & 96.96\% & 58.09\%    \\ \cline{2-5}
                           & 4\% & 32 & 96.08\% & 52.38\%     \\ \cline{2-5}
                           & 5\% & 40 & 95.32\% & 56.19\%    \\ \hline
\end{tabular}
}
\end{table}

\begin{figure}[ht!]
    \centering
    \includegraphics[width = 0.9\columnwidth]{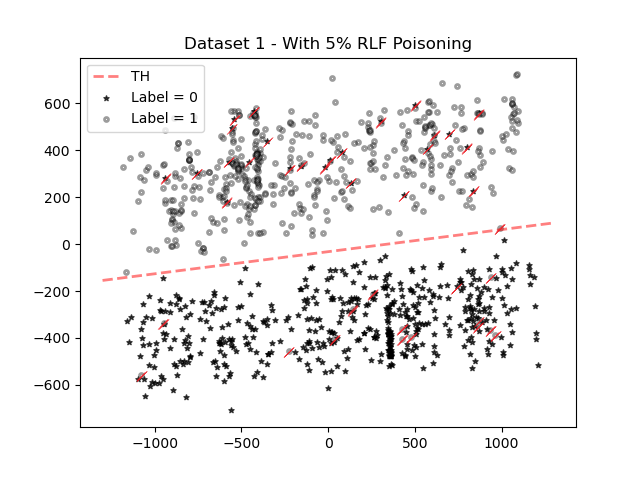}
    \caption{Illustration of 5\% samples poisoned in Dataset 1
}
    \label{fig:Poisoned Dataset 1}
\end{figure}

To get deeper insights, we generated plots of the poisoned datasets to visualize the manipulated labels. A threshold in red that acts as a decision margin is introduced to differentiate between benign URLs labeled as ``0'' and malicious URLs labeled as ``1''. 
Figure \ref{fig:Poisoned Dataset 1} displays Dataset 1 after randomly flipping 5\% of the labels, applied on both benign and malicious URLs. The 40 manipulated samples from the 5\% poisoned data are observed in this figure as they are inaccurately positioned with reference to the decision margin. We crossed in red the 40 samples in Figure \ref{fig:Poisoned Dataset 1} for clear observation.

\subsection{Apply Defense on RF}

We analyzed the effectiveness of our defense mechanism against random LF attacks on RF classifiers. This defense mechanism is based on the $K$-NN approach that predicts true labels by identifying the $K$ nearest labels from the original dataset $D$ to each label from the untrusted dataset $D'$. Initially, we started with identifying the best value of \textit{K} for each dataset to ensure a minimum error rate in our results using original datasets. Table \ref{table:Kvalue} presents the best value of \textit{K} for each dataset obtained using Algorithm \ref{alg:K} and we can observe that $K$ value can vary for different datasets. For instance, $K$ is set to "33" for Dataset 1 while for Dataset 2, $K$ is set to "5".

\begin{table}[H]
\centering
\caption{The Best value of \textit{K} for all six datasets}
\label{table:Kvalue}
\small 
\begin{tabular}{|l|l|}
\hline
Original Dataset & Value of \textit{$K$} \\
\hline
\textbf{Original Dataset 1} & 33 \\ \hline
\textbf{Original Dataset 2} & 5 \\ \hline
\textbf{Original Dataset 3} & 3 \\ \hline
\textbf{Original Dataset 4} & 3 \\ \hline
\textbf{Original Dataset 5} & 3 \\ \hline
\textbf{Original Dataset 6} & 3 \\ \hline

\hline
\end{tabular}
\end{table}

Using the \textit{K} values, we applied our defense mechanism with the aim to detect the poisoned labels, activate an alarm for each poisoned label, and recovering the value of the poisoned label. After applying the defense mechanism on the entire dataset, this will provide a full recovered dataset from the applied random LF attack.

\begin{table}[H]
\centering
\caption{Accuracy of RF model on the six datasets with K-NN defense against Random LF attack}
\label{table:Defense-on-all}
\resizebox{\linewidth}{!}{%
\begin{tabular}{|l|l|l|l|} 
\hline
Dataset & Poison\% & Tr. Accuracy~~ & Det. Poisoned 
labels \\ \hline
\multirow{4}{*}{\textbf{Dataset 1}} & 2\%  & 99.87\%    & 17 \\ \cline{2-4}
                           & 3\%  & 99.87\%    & 23 \\ \cline{2-4}
                           & 4\%  & 100.00\%   & 31 \\ \cline{2-4}
                           & 5\%  & 100.00\%   & 40  \\ \hline
\multirow{4}{*}{\textbf{Dataset 2}} & 2\%  & 99.87\%    & 18  \\ \cline{2-4}
                           & 3\%  & 100.00\%   & 24  \\ \cline{2-4}
                           & 4\%  & 100.00\%   & 34  \\ \cline{2-4}
                           & 5\%  & 99.87\%    & 40  \\ \hline
\multirow{4}{*}{\textbf{Dataset 3}} & 2\%  & 100.00\%   & 18  \\ \cline{2-4}
                           & 3\%  & 100.00\%   & 24  \\ \cline{2-4}
                           & 4\%  & 100.00\%   & 31  \\ \cline{2-4}
                           & 5\%  & 100.00\%   & 40   \\ \hline
\multirow{4}{*}{\textbf{Dataset 4}} & 2\%  & 100.00\%   & 16   \\ \cline{2-4}
                           & 3\%  & 100.00\%   & 20   \\ \cline{2-4}
                           & 4\%  & 100.00\%   & 30   \\ \cline{2-4}
                           & 5\%  & 100.00\%   & 38   \\ \hline
\multirow{4}{*}{\textbf{Dataset 5}} & 2\%  & 100.00\%   & 18   \\ \cline{2-4}
                           & 3\%  & 100.00\%   & 26    \\ \cline{2-4}
                           & 4\%  & 99.89\%    & 34    \\ \cline{2-4}
                           & 5\%  & 100.00\%   & 42    \\ \hline
\multirow{4}{*}{\textbf{Dataset 6}} & 2\%  & 100.00\%   & 16    \\ \cline{2-4}
                           & 3\%  & 100.00\%   & 24    \\ \cline{2-4}
                           & 4\%  & 100.00\%   & 32     \\ \cline{2-4}
                           & 5\%  & 99.87\%    & 38    \\ \hline
\end{tabular}
}
\end{table}

As per the results in Table \ref{table:Defense-on-all}, our mechanism successfully recovered the majority of the poisoned labels, and the training accuracy has increased. For instance, with 5\% poison rate, Dataset 3 training accuracy increased after applying the defense from 95.19\% to 100\% by recovering 40 detected poisoned labels while Dataset 2 training accuracy increased from 95.18\% to 99.87\% accuracy by recovering 40 detected poisoned labels. This proves the effectiveness of our mechanism in detecting poisoned labels and correcting them to their true label. Taking into account that achieving 100\% accuracy in most cases may sound unrealistic, however, this is due to repeatedly training the RF model on the same dataset.

\begin{figure}[ht!]
    \centering
    \includegraphics[width = 1.0\columnwidth]{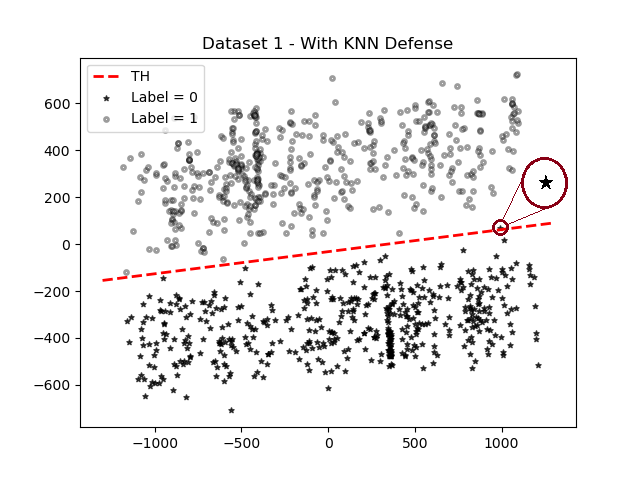}
    \caption{Recovered Dataset 1 from 5\% poisoned samples
    }
    \label{fig:Recovered Dataset 1}
\end{figure}

For the final analysis, we also generated plots for the recovered datasets obtained after applying the defense mechanism. By comparing the poisoned dataset plot and the recovered dataset plot, we can clearly visualize the effectiveness of our approach. Figure \ref{fig:Recovered Dataset 1} displays Dataset 1, which was initially manipulated by 5\% of its labels, that is recovered by our defense mechanism, and all labels are split correctly with reference to the red decision margin. If we compared Figure\ref{fig:Recovered Dataset 1} with the original dataset in Figure\ref{fig:CleanRun Dataset 1}, we can observe a slight difference between the two figures where a label on the decision margin is denoted as ``0'' in Figure \ref{fig:Recovered Dataset 1} however the same label is denoted as ``1'' in Figure \ref{fig:CleanRun Dataset 1}.

\subsection{Discussion}

We initially started our experiment by running the RF model on original datasets to observe the behavior of the model and 100\% training accuracy rate is obtained for five datasets. We set a reference point to measure the impact of the random LF attack and the efficiency of our proposed defense method.
Our findings underscored the possibility of an attacker manipulating datasets on RF models through label poisoning, as evidenced by manipulation, a minor degradation is observed in model accuracy between 95-98\% is usually negligible in ML systems in addition to achieving ASR in the range of 50.47- 64.76\%. 
Finally, the application of the \textit{K}-NN defense method demonstrated a way to mitigate the impact of label poisoning, with notable improvements in the model's robustness through detecting poisoned labels and achieving an accuracy rate in the range of 99.87-100\% with the recovered datasets. Taking into account, the vital role of having an original dataset to implement our defense method.


%

%

\section{Conclusions and Future works}
\label{Conclusions}
\label{sec:conc-section}

In this paper, we shed light on several crucial aspects of random LF attacks against malicious URL detectors and possible defense strategies through the results of our analysis. Initially, a LF attack is applied through random flipping of benign and malicious labels at a small poisoning rate of 2-5\%. According to the results of our study, corrupting the samples within the training dataset during the training phase and achieving ASR between 50-65\% within 2-5\% poisoning rate makes this attack a more convenient option for an attacker who aims to impact the performance of a model. It is important to note that the LF attack succeeded in fooling the detector and playing the role of a ghost. Accordingly, the implementation of a defense model is crucial to overcome the impact of LF attacks and recover the poisoned datasets. Our defense method has successfully detected the poisoned labels, predicted true labels, and improved the accuracy of the model up to 100\%.

Our study has the potential to function as a valuable tool for appraising the robustness of RF models and improving their performance against malicious URLs. The focus of future research will counter other types of backdoor attacks in URL detectors that use the RF classifier to implement possible attack recognition and defense mechanisms. The main objective of these mechanisms will be to identify the existence of backdoors within the training dataset by analyzing the distribution of features in the samples. 
Further investigations can play a pivotal role in advancing not only our study but also the wider domain of ML security.

%

\bibliographystyle{IEEEtran}
\bibliography{References}

\vskip -3\baselineskip plus -1fil
\begin{IEEEbiography}
[{\includegraphics[width=1in,height=1.25in]{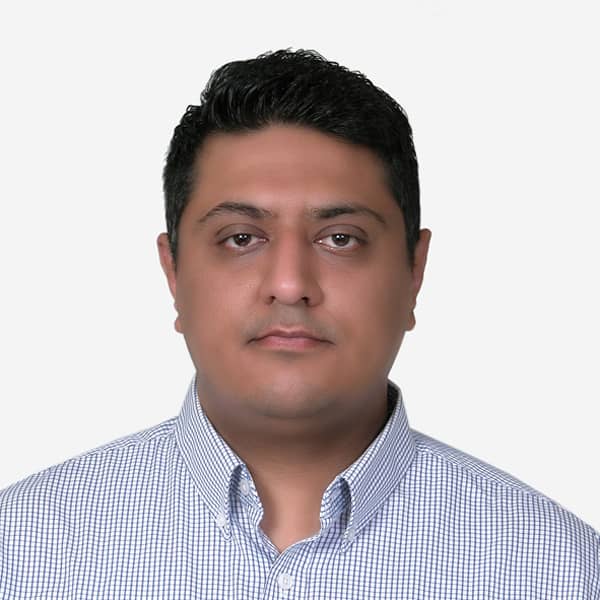}}]
{Ehsan Nowroozi} (Senior IEEE Member) is a Senior Lecturer (Associate Professor) at Ravensbourne University London, Department of Business and Computing, London, UK. He received his doctorate from Siena University in 2020. Ehsan’s work emphasizes determining and defending against adversarial threats. His research has had a significant impact on the development of AI-cyber and the ability to defend against cyber threats. He had four postdocs in different high-prestige universities, including a Research Fellow at the Centre for Secure Information Technologies (CSIT) at Queen’s University Belfast in the United Kingdom, a Research Fellow at the Security and Privacy Research Group (SPRITZ) at the University of Padua in Italy, a Research Fellow at the Visual Information Processing and Protection (VIPP) at the University of Siena in Italy, a Research Fellow at the Sabanci University in Turkey. He was also an assistant professor at Bahcesehir University, Istanbul, Turkey. He has worked on a variety of projects funded by renowned institutions, such as DARPA, the Air Force Research Laboratory (AFRL) of the U.S. government, the Italian Ministry of University and Research (MUR), and THALES United Kingdom. He serves as a reviewer for prominent journals, such as IEEE TNSM, IEEE TIFS, and IEEE TNNLS. In addition, he has been a senior member of the Institute of Electrical and Electronics Engineers (IEEE) since 2022 and an ACM Professional member since 2023.
  \\ 

\end{IEEEbiography}

\vskip -3.6\baselineskip plus -1fil
\begin{IEEEbiography}
[{\includegraphics[width=0.9in,height=1.25in]{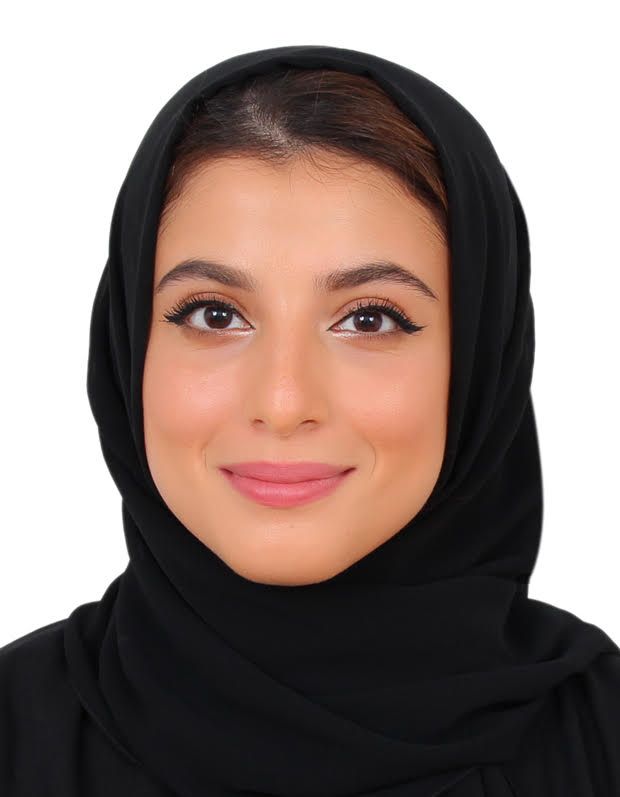}}]
{Nada Jadalla} received her Master's from Bahcesehir University. She received a bachelor’s degree in electrical/telecommunication engineering from Ajman
University, Ajman, UAE in 2017. Her main research interest is in the area of Machine Learning,
Artificial Intelligence and Cybersecurity. She worked
in the technology and regulatory field within the telecommunication industry and she is also a student member of IEEE institution.  \\ 

\end{IEEEbiography}

\vskip -3.6\baselineskip plus -1fil
\begin{IEEEbiography}
[{\includegraphics[width=0.9in,height=1.15in]{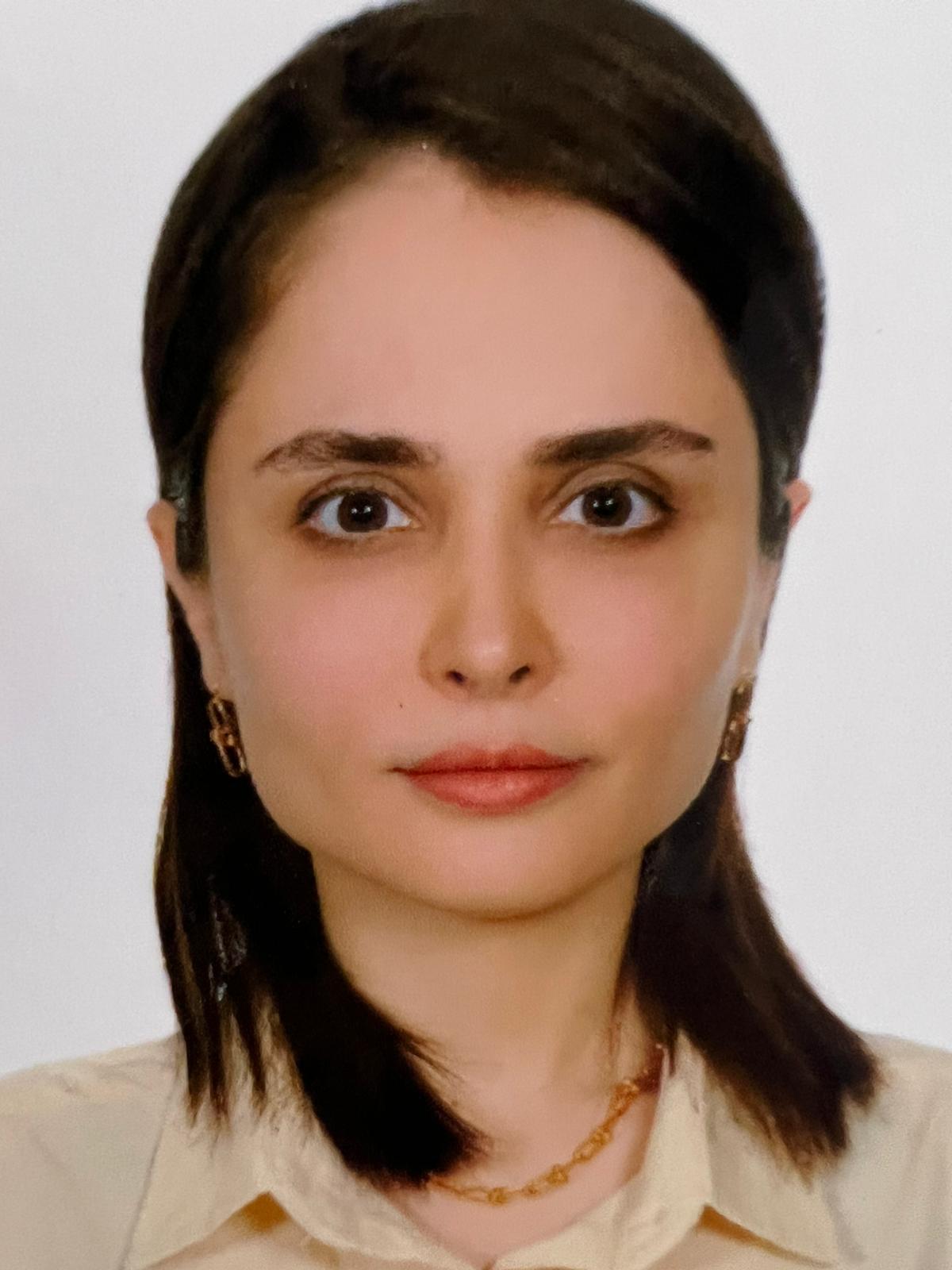}}]
{Samaneh Ghelichkhani} received a Master's degree in Advanced Computer Science from the University of Leeds, United Kingdom, and in Information Technology Engineering (Networking branch) from Islamic Azad University. Furthermore received a Bachelor’s degree in Information Technology Engineering from Islamic Azad University, Iran.  Her main research interest is in artificial intelligence including machine and deep learning, and networks.   \\ 

\end{IEEEbiography}

\vskip -5\baselineskip plus -1fil
\begin{IEEEbiography}
[{\includegraphics[width=0.9in,height=1.15in]{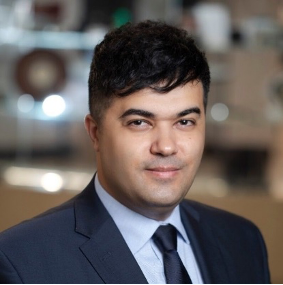}}]
{Alireza Jolfaei} is an Associate Professor in Cybersecurity and Networking at the College of Science and Engineering at Flinders University. He is a Senior Member of the IEEE and a Distinguished Speaker of the ACM on the topic of Cybersecurity. He has previously been a faculty member with Macquarie University, and Federation University in Australia, and Temple University in the USA. He received a Ph.D. degree in Applied Cryptography from Griffith University, Gold Coast, Australia. His main research interest is in Cyber-Physical Systems Security, where he investigates the hidden interdependencies in industrial communication protocols and aims to provide fundamentally new methods for security-aware modeling, analysis, and design of safety-critical cyber-physical systems in the presence of cyber-adversaries. He has been a chief investigator of several internal and external grants. He successfully supervised eight HDR students to completion. He received the prestigious IEEE Australian Council award for his research paper published in the IEEE Transactions on Information Forensics and Security.   \\ 

\end{IEEEbiography}

\end{document}